\documentclass[twocolumn,showpacs,amsmath,amssymb,floatfix]{revtex4}
\usepackage{graphicx}

\begin{document}

\title{Magnetoresistance of atomic-scale electromigrated nickel
nanocontacts}

\author{Z.K. Keane, L.H. Yu, and D. Natelson}

\affiliation{Department of Physics and Astronomy, Rice University, 6100
Main St., Houston, TX 77005}

\date{\today}

\begin{abstract}

We report measurements of the electron transport through atomic-scale
constrictions and tunnel junctions between ferromagnetic electrodes.
Structures are fabricated using a combination of e-beam lithography
and controlled electromigration. Sample geometries are chosen to allow
independent control of electrode bulk magnetizations.  As junction size is
decreased to the single channel limit, conventional anisotropic
magnetoresistance (AMR) increases in magnitude, approaching the size
expected for tunneling magnetoresistance (TMR) upon tunnel junction
formation.  Significant mesoscopic variations are seen in the
magnitude and sign of the magnetoresistance, and no evidence is
found of large ballistic magnetoresistance effects.

\end{abstract}
\maketitle

Magnetoresistive effects in ferromagnetic structures are of
considerable technological and scientific interest.  The anisotropic
magnetoresistance (AMR) results from spin-orbit scattering and is
manifested as a change in resistivity, $\rho$, as a function of the
relative orientation of the magnetization ${\bf M}$ and the current
density ${\bf J}$.  In Ni, $(\rho(J || M)-\rho(J \perp M))/\rho(J ||
M) \approx 0.02$.  Tunneling magnetoresistance (TMR) results from the
difference in majority and minority densities of states at the Fermi
level.  Tunneling resistance, $R$, is generally enhanced for
antialigned magnetizations of the electrodes on either side of the
tunnel barrier.  The magnitude of TMR in large area junctions is
$\equiv (R_{\uparrow \downarrow}-R_{\uparrow \uparrow})/R_{\uparrow
\uparrow} = 2P^{2}/(1-P^{2})$, where $P$ is the spin polarization at
the Fermi level\cite{Julliere55PLA}.

The magnetoresistance of atomic-scale constrictions in magnetic wires
has been the subject of intense interest since the initial report of
large ballistic magnetoresistance (BMR) in junctions between Ni
wires\cite{GarciaetAl99PRL}.  Reports of BMR magnitudes far in excess
of typical AMR and TMR effects have generated considerable
controversy, including concerns about magnetostrictive
artifacts\cite{EgelhoffetAl04JAP}.  It is therefore of much interest
to examine constrictions fabricated in a geometry that minimizes these
effects and allows temperature-dependent studies of junction
magnetoresistances.  Recent experiments along these lines have used
mechanical break junctions\cite{ViretetAl02PRB}, planar
electrochemically grown junctions\cite{YangetAl04APL}, ballistic
nanopores\cite{OzatayetAl04JAP}, and ion-beam-formed
constrictions\cite{MonteroetAl04PRB}

In this Letter we report measurements of the magnetoresistance through
few-atom and single-atom contacts between planar Ni electrodes, as
well as planar Ni-Ni tunnel junctions.  Junctions are fabricated by a
combination of electron beam lithography and controlled
electromigration.  This allows the examination of individual
nanostructures with junction configurations serially modified from
planar films to few-atom contacts to vacuum tunnel junctions.  These
planar structures are chosen to minimize magnetostrictive effects, as
discussed below.  Small junction size is confirmed by evidence of
conductance quantization and discrete switching.  At 10 K,
conventional AMR is observed in large junctions, and increases in
magnitude as the number of channels approaches one.  We observe
significant sample-to-sample variation in the shape and sign of the
magnetoresistance, with an upper limit on the magnitude consistent
with TMR in Ni of known properties.  This variability, typical of
mesoscopic systems, suggests that the bulk magnetization of the
electrodes is not simply related to the local magnetization of the few
atoms directly relevant for tunneling.

\begin{figure}[h!]
\begin{center}
\includegraphics[clip, width=8cm]{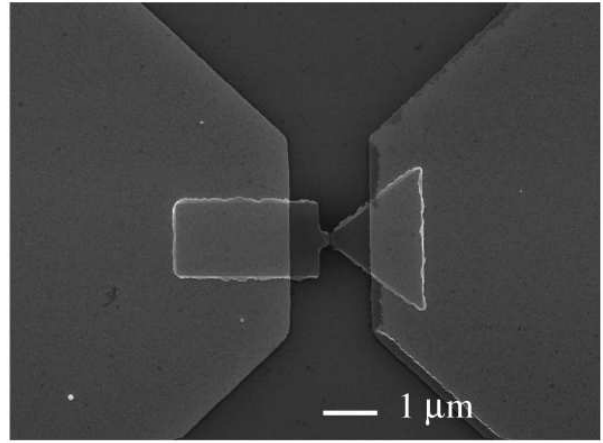}
\end{center}
\vspace{-3mm}
\caption{\small Scanning electron micrograph of a typical device.
Leads are 1~nm Ti/30~nm Au; constricted wire is 20~nm thick Ni. }
\label{fig1}
\vspace{-3mm}
\end{figure}

Devices are fabricated by a two-step lithography process on test grade
$p+$ Si wafers coated with 200 nm thermal oxide.  Ni structures are
defined by e-beam lithography and e-beam evaporation (20 nm thick Ni
film deposited at 2 \AA/s in a system with $\sim 10^{-7}$~mB base
pressure).  This is followed by a second lithography step, Ar ion
sputter cleaning to ensure good contact to the Ni layer, and
evaporation of 1 nm Ti/30 nm Au leads to make electrical contact.
Sample geometries, shown in Fig.~\ref{fig1}, were chosen to minimize
magnetostrictive effects by anchoring the bulk electrodes firmly to
the substrate, and to create a well-defined domain structure near the
constriction so that the data could be more easily interpreted.  The
micron scale of the Ni pads increases the likelihood that each will
consist of a single domain. The electrode shapes favor controlled
relative reorientation of their bulk
magnetizations\cite{JedemaetAl01Nature,PasupathyetAl04Science}.  In
the absence of an external magnetic field, ${\bf M}$ is favored to lie
in the plane of the electrodes and parallel to the current.

The Ni constrictions are progressively broken by
electromigration\cite{ParketAl99APL} to achieve contacts ranging from
a few atoms to vacuum tunnel junctions.  All measurements are
performed at 10~K in a variable temperature vacuum cryostat to
mitigate oxidation of the Ni atoms near the contact.  The system is
relatively stable at this temperature, allowing measurement of the
same device in multiple configurations; in this way we were able to
observe the evolution of these devices from bulk metal through the
ballistic regime into the tunneling regime.

\begin{figure}[h!]
\begin{center}
\includegraphics[clip, width=8cm]{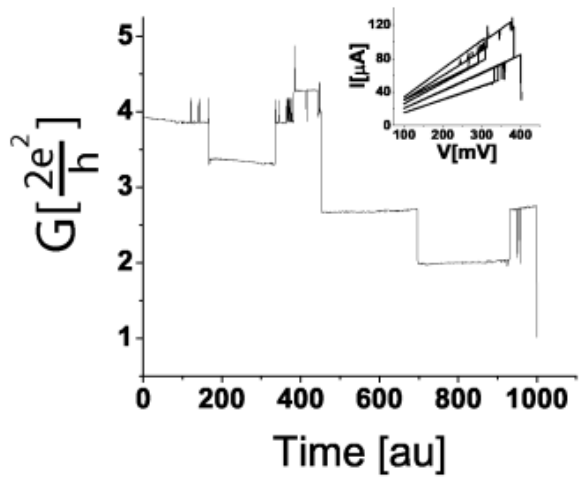}
\end{center}
\vspace{-3mm}
\caption{\small Conductance vs. time for a successful electromigration
run, demonstrating discontinuities and conductance plateaux.  Total
duration of the data shown is approximately 10 seconds. (inset) $I-V$
curves representing the same data.  }
\label{fig2}
\vspace{-3mm}
\end{figure}

To achieve few-atom contacts and clean vacuum tunnel junctions
reliably, precise control of the electromigration procedure is
paramount; our procedure for nickel closely follows that previously
reported for room-temperature gold nanojunctions\cite{StrachanetAl05APL}.
Achieving stable few-channel junctions is extremely challenging, with
a success rate of $\sim$ 8\%.  Current-voltage curves and corresponding
conductance data from a successful electromigration run resulting in a
13~k$\Omega$ device are shown in Fig.~\ref{fig2}.  The discontinuities
in the last few $I-V$ curves, and the corresponding conductance
plateaux, indicate that the device under test likely consists of a few
atoms at its narrowest point.  Such few-channel junctions often
exhibit telegraph noise with conductance changes $\sim e^{2}/h$, also
consistent with extremely narrow constrictions.

\begin{figure}[h!]
\begin{center}
\includegraphics[clip, width=8cm]{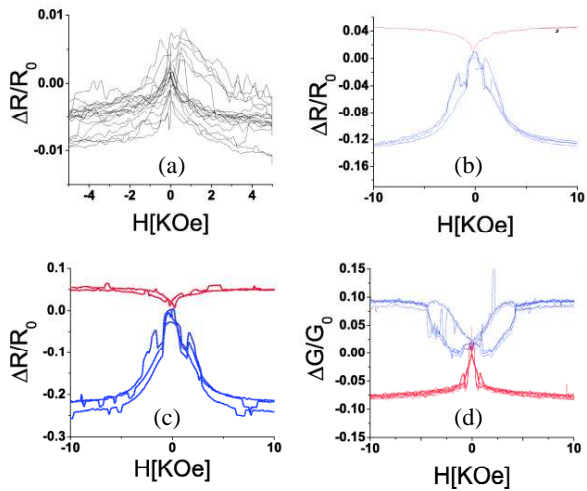}
\end{center}
\vspace{-3mm}
\caption{\small a) Anisotropic magnetoresistance of a typical device (sample E)
before electromigration.  b) In-plane (upper, red online) and
out-of-plane (lower, blue online) magnetoresistance of sample E at 5.8~k$\Omega$.  c) In-plane (upper, red online) and out-of-plane (lower, blue
online) magnetoresistance of sample E further broken to
13~k$\Omega$.  d) In-plane (lower, red online) and out-of-plane
(upper, blue online) magnetoconductance of a 10~M$\Omega$ tunnel
junction (sample C).}
\label{fig3}
\vspace{-3mm}
\end{figure}

A family of magnetoresistance curves from one of these devices is
shown in Fig.~\ref{fig3}.  The magnetization of the leads lies in the
plane of the film until a coercive external field is applied out of
plane.  An {\it in situ} rotation stage allows the acquisition of
magnetoresistance data as a function of field orientation for a single
junction configuration.  The magnetoresistance under a transverse
field evolves gradually from a small AMR in
the bulk to a larger, similarly-shaped curve as the device is broken
into the ballistic regime.  The initial magnitude of the out-of-plane
MR in an unbroken starting device is typically a few tenths of a
percent; while AMR in nickel films is typically ~2\%, the smaller
value is reasonable since initially the measured resistance is
dominated by the leads and wiring.  As the device is progressively
broken, the MR quickly approaches and then surpasses
the expected magnitude for bulk AMR: in few- and single-channel
devices, this effect can approach 20\%.  Finally, in the tunneling
regime, we see fairly typical MR magnitudes for a vacuum tunnel
junction, with TMR values of 10-20\%.  Table~\ref{tab1} shows several samples
measured in different electromigrated configurations at 10~K.

\begin{table}
\caption{\small Magnetoresistance of samples under various electromigrated configurations.}
\begin{tabular}{c c c c}
\hline 
\hline
Sample & Resistance & Longitudinal MR [\%] & Transverse MR [\%]\\ 
\hline
A & 7.1 k$\Omega$ & 0.78 & 1.14\\
A & 83 k$\Omega$ & 20.7 & 16.4\\
B & 13 k$\Omega$ & 20.5 & 8.03\\
B & 5 M$\Omega$ & 11.3 & 10.7\\
C & 10 M$\Omega$ & 9.43 & 7.07\\
D & 13 k$\Omega$ & 5.13 & 10.7\\
D & 200 k$\Omega$ & 13.3 & [not measured]\\
E & 5.8 k$\Omega$ & 3.3 & 15.4\\
E & 13 k$\Omega$ & 8.69 & 21.9\\
\hline
\hline
\end{tabular}
\label{tab1}
\end{table}

A closer look at the curves in Fig.~\ref{fig3} reveals some aspects,
other than the surprisingly large magnitude fo the AMR-like effect, in
which the behavior of these devices diverges from traditional AMR.
The most readily apparent of these unusual behaviors is the appearance
of switching features at applied fields of around 2 KOe perpendicular
to the current.  These features are not observed in any devices prior
to electromigration.  The hysteretic nature of these switching
features suggests that they may be due to domain reversal in the Ni
metal.  In both in-plane and out-of-plane field sweeps, the magnitude
and sign of the TMR has significant variability from device to device.

Magnetostriction is a possible confounding effect: one need only cause
a single- atom mechanical contact to shift by a fraction of a
nanometer to cause a large conductance change.  A recent
experiment\cite{GabureacetAl04PRB} designed to quantify the effects of
magnetostriction in nickel mechanical break junctions indicates that,
in a geometry with a 650~nm undercut beneath the bridge,
magnetostriction accounted for a 40\% change in the resistance of an
atomic-scale contact.  Magnetostriction is very unlikely to account
for our results for two reasons.  First, in our geometry, the Ni wires
are constrained by the silicon substrate everywhere except for the
immediate neighborhood of the constriction; the length of the bridge
which is unconstrained by the substrate below is at most $\sim$10~nm,
which should result in a much smaller magnetostrictive effect.
Second, the magnetoresistance curves shown in \cite{GabureacetAl04PRB}
are qualitatively different from those reported here, particularly in
the lack of fine structure.

The complicated MR patterns and variability indicate that domain
structure in the bulk electrodes is not simply related to the
atomic-scale magnetization at few-atom contact or point of tunneling.
Since there is no evidence of Coulomb blockade as these devices
approach the TMR limit, it is unlikely that the effects seen are a
result of unintentional nanoparticle formation during
electromigration.  Similar variability and complicated magnetic
structure has also been seen in mechanical break junction experiments
in the few-channel regime\cite{ViretetAl02PRB}.   Such strong
sensitivity to detailed contact geometry has also been supported
theoretically\cite{JacobetAl05PRB}.

Three factors are likely to be relevant to understanding these
observations.  First, single-molecule transistor measurements with
ferromagnetic leads\cite{PasupathyetAl04Science} have explicitly
demonstrated that effective exchange fields at surface atoms can be
large (70 T) and different from the bulk.  Second, tunneling via
localized states (possibly surface states in this case) has been
demonstrated to lead to inverted TMR\cite{TsymbaletAl03PRL}, as
have highly transmitting channels\cite{Kim05PRB}.  Third,
it is possible that trace amounts of NiO$_{x}$ or unintended
adsorbates at the tunneling point can cause local perturbations of the
tunneling spins.  Detailed atomic-scale variations in the junctions
clearly can have a profound influence on relevant magnetoresistive
processes.

In summary, in nanoscale Ni junctions we observe an evolution of
magnetoresistance from ordinary AMR in wide junctions, to an enhanced
AMR in few-channel wires, to TMR in tunnel junctions, with large
sample-to-sample variability in the shapes and signs of the TMR.  No
magnetoresistances are observed that are larger than those expected
from the known polarization of Ni.  The mesoscopic variation in MR
indicates that the local junction environment can have a strong affect
on the spin of the tunneling carriers.  Further study is required
to determine the precise physics behind this effect.

We note that Bolotin, Kuemmeth, Pasupathy, and Ralph have recently
posted independent results of a similar experiment\cite{BolotinetAl05}.

This work was supported by NSF NER award ECS-0403457, the 
David and Lucille Packard Foundation, and an Alfred P. Sloan Foundation
research fellowship.


\begin{thebibliography}{10}

\bibitem{Julliere55PLA}
M. Julliere, Phys. Lett. A {\bf 54}, 225 (1955).

\bibitem{GarciaetAl99PRL}
N. Garc{\'i}a, M. Mu{\~n}oz, and Y.-W. Zhao, Phys. Rev. Lett. {\bf 82}, 2923 (1999).

\bibitem{EgelhoffetAl04JAP}
W.F. Egelhoff, Jr., L. Gan, H. Ettedgui, Y. Kadmon, C.J. Powell, P.J. Chen, A.J. Shapiro, R.D. McMichael, J.J. Mallett, T.P. Moffat, M.D. Stiles, E.B. Svedberg, J. Appl. Phys. {\bf 95}, 7554 (2004).

\bibitem{ViretetAl02PRB}
M. Viret, S. Berger, M. Gabureac, F. Ott, D. Olligs, I. Petej, J.F. Gregg, C. Fermon, G. Francinet, and G. Le Goff, Phys. Rev. B {\bf 66}, 220401(R) (2002).

\bibitem{YangetAl04APL}
C.-S. Yang, C. Zhang, J. Redepenning, and B. Doudin, Appl. Phys. Lett. {\bf 84}, 2865 (2004).

\bibitem{OzatayetAl04JAP}
O. Ozatay, P. Chalsani, N.C. Emley, IN. Krivorotov, and R.A. Buhrman, J. Appl. Phys. {\bf 95}, 7315 (2004).


\bibitem{MonteroetAl04PRB}
M.I. Montero, R.K. Dumas, G. Liu, M. Viret, O.M. Stoll, W.A.A. Macedo, and I.K. Schuller, Phys. Rev. B {\bf 70}, 184418 (2004).

\bibitem{JedemaetAl01Nature}
F.J. Jedema, A.T. Filip, and B.J. van Wees, Nature {\bf 410}, 345 (2001).

\bibitem{PasupathyetAl04Science}
A.N. Pasupathy, R.C. Bialczak, J. Martinek, J.E. Grose, L.A.K. Donev, P.L. McEuen, and D.C. Ralph, Science {\bf 306}, 86 (2004).

\bibitem{ParketAl99APL}
H. Park, A.K.L. Lim, A.P. Alivisatos, J. Park, and P.L. McEuen, Appl. Phys. Lett. {\bf 75}, 301 (1999). 

\bibitem{StrachanetAl05APL}
D.R. Strachan, D.E. Smith, D.E. Johnston, T.-H. Park, M.J. Therien, D.A. Bonnell, and A.T. Johnson, Appl. Phys. Lett. {\bf 86}, 043109 (2005).

\bibitem{GabureacetAl04PRB}
M. Gabureac, M. Viret, F. Ott, and C. Fermon, Phys. Rev. B {\bf 69}, 100401 (2004).

\bibitem{JacobetAl05PRB}
D. Jacob, J. Fern{\'a}ndez-Rossier, and J.J. Palacios, Phys. Rev. B {\bf 71}, 220403(R) (2005).


\bibitem{TsymbaletAl03PRL}
E.Y. Tsymbal, A. Sokolov, I.F. Sabirianov, and B. Doudin, Phys. Rev. Lett. {\bf 90}, 186602 (2003).

\bibitem{Kim05PRB}
T.-S. Kim, Phys. Rev. B {\bf 72}, 024401 (2005).

\bibitem{BolotinetAl05}
K.I. Bolotin, F. Kuemmeth, A.N. Pasupathy, and D.C. Ralph, cond-mat/0510410 (2005).

\end{thebibliography}
\end{document}